\documentclass[cameraready]{Interspeech}

\title{AcoustEmo: Open-Vocabulary Emotion Reasoning via Utterance-Aware Acoustic Q-Former}

\author[affiliation={1}, orcid=0000-0003-3174-6588]{Liyun}{Zhang}
\author[affiliation={2}, orcid=0009-0006-2402-0075]{Xuanmeng}{Sha}
\author[affiliation={2}, orcid=0000-0003-1501-9719]{Shuqiong}{Wu}
\author[affiliation={2}, orcid=0009-0002-1904-767X]{Fengkai}{Liu}

\address{
    $^1$ The University of Tokyo, Tokyo, Japan \\
    $^2$ The University of Osaka, Osaka, Japan
}

\email{liyun.zhang@lab.ime.cmc.osaka-u.ac.jp}


\usepackage{comment}

\begin{document}

\maketitle

\begin{abstract}
    Multimodal Large Language Models (MLLMs) excel in Open-Vocabulary (OV) emotion recognition but often neglect fine-grained acoustic modeling. Existing methods typically use global audio encoders, failing to capture subtle, local temporal dynamics like micro-prosody and intonation shifts within individual utterances. To address this, we propose AcoustEmo, a time-sensitive MLLM featuring a novel Utterance-Aware Acoustic Q-Former. Our approach utilizes a timestamp-synchronized sliding window to dynamically extract segment-level audio tokens instead of coarse global representations. This enables the model to explicitly trace the temporal evolution of subtle acoustic clues and capture deep contextual dependencies in dialogues. Experiments on the Explainable Multimodal Emotion Recognition (EMER) task show that AcoustEmo significantly enhances complex emotion reasoning, outperforming baselines while maintaining robust contextual accuracy.
\end{abstract}

\section{Introduction}
Multimodal Large Language Models (MLLMs) have shown exceptional capabilities in integrating visual, acoustic, and linguistic modalities for complex video understanding tasks \cite{Video-llama}. The integration of these modalities enables machines to approach human-like perception, which is crucial for downstream applications such as empathetic conversational agents, mental health monitoring, and advanced human-computer interaction \cite{Panoptic-tcsvt, Panoptic-wacv, Panoptic1, Thermal-to-Color, PhD, 3DFacePolicy, 3DGesPolicy, uneven, Momentum, Supplementary}. Recently, adapting MLLMs for Open-Vocabulary (OV) emotion recognition---such as the Explainable Multimodal Emotion Reasoning (EMER) task \cite{EMER(Multi)}---has attracted significant attention. This task requires the model not only to fuse multimodal information but also to conduct deep reasoning to capture highly sensitive and dynamically changing emotional states.

Despite their remarkable performance, existing MLLMs exhibit a critical limitation in their acoustic modeling paradigms. Most methods rely on a simplistic global audio encoder, where the entire audio track of a video is compressed into coarse, global sequence tokens. While efficient, this global processing paradigm intrinsically fails to capture local, fine-grained temporal dynamics of speech. Emotion in speech is heavily conveyed through subtle, fleeting acoustic clues---such as micro-prosody, sudden intonation shifts, trembling, and speech rate variations---that occur within specific spoken utterances. Neglecting these utterance-level acoustic dynamics restricts the model's expected effectiveness in nuanced emotion reasoning. Previous works, such as MicroEmo \cite{zhang2024microemo}, have successfully attempted to capture subtle temporal dynamics in facial visual features, yet the critical temporal dynamics within the acoustic modality remain largely underexplored.

To this end, we propose \textbf{AcoustEmo}, a novel time-sensitive MLLM architecture. To the best of our knowledge, this is the first work to direct attention toward local acoustic dynamics and the contextual dependencies of utterance-aware audio segments in open-vocabulary emotion reasoning. AcoustEmo replaces the conventional global audio encoder with a novel Utterance-Aware Acoustic Q-Former. By utilizing a timestamp-synchronized sliding window, our module dynamically extracts fine-grained, segment-level acoustic tokens that strictly align with textual utterances. 

The main contributions of this work are summarized as follows:
\begin{itemize}
    \item We propose a novel Utterance-Aware Acoustic Q-Former tailored for MLLMs. Overcoming the limitations of global audio encoders, this module dynamically extracts segment-level acoustic tokens to explicitly capture local temporal dynamics and subtle acoustic clues.
    \item We introduce a timestamp-synchronized acoustic sliding window mechanism. By strictly aligning local acoustic feature sequences with utterance-level transcription timestamps, our approach effectively models deep contextual dependencies in continuous dialogues.
    \item We empirically demonstrate the superiority of fine-grained acoustic modeling on the EMER task. Extensive experiments reveal that our time-sensitive architecture significantly enhances open-vocabulary emotion reasoning.
\end{itemize}

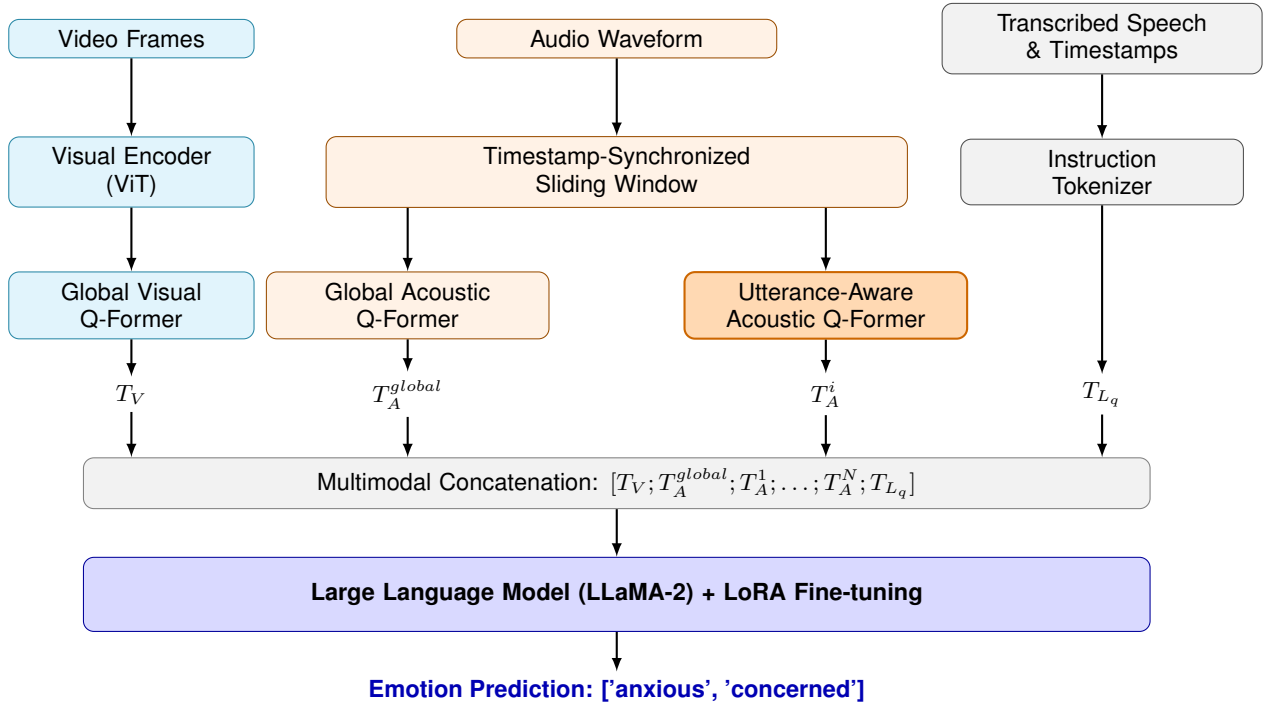
\begin{figure*}[t]
\centering
\resizebox{0.98\linewidth}{!}{
\begin{tikzpicture}[
    >=latex,
    box/.style={draw, rounded corners, align=center, font=\sffamily\small, inner sep=1ex},
    vbox/.style={box, fill=cyan!10, draw=cyan!60!black, text width=3cm},
    abox/.style={box, fill=orange!10, draw=orange!60!black},
    tbox/.style={box, fill=gray!10, draw=gray!60!black},
    corebox/.style={box, fill=orange!30, draw=orange!80!black, thick, text width=3.5cm},
    llmbox/.style={box, fill=blue!15, draw=blue!60!black, text width=14cm, minimum height=1cm, font=\sffamily\bfseries},
    concatbox/.style={box, fill=black!5, draw=black!50, text width=14cm, minimum height=0.6cm, font=\sffamily},
    token/.style={font=\sffamily\bfseries}
]

\node[vbox] (v_in) at (-6.5, 0) {Video Frames};
\node[vbox] (v_enc) at (-6.5, -1.8) {Visual Encoder\\(ViT)};
\node[vbox] (v_qf) at (-6.5, -3.6) {Global Visual\\Q-Former};
\node[token] (t_v) at (-6.5, -4.8) {$T_V$};

\node[abox, text width=4cm] (a_in) at (0, 0) {Audio Waveform};
\node[abox, text width=7.5cm] (a_slide) at (0, -1.8) {Timestamp-Synchronized\\Sliding Window};

\node[abox, text width=3.5cm] (a_glob) at (-2.8, -3.6) {Global Acoustic\\Q-Former};
\node[corebox] (a_loc) at (2.8, -3.6) {Utterance-Aware\\Acoustic Q-Former};

\node[token] (t_a_g) at (-2.8, -4.8) {$T_A^{global}$};
\node[token] (t_a_l) at (2.8, -4.8) {$T_A^i$};

\node[tbox, text width=4cm] (t_in) at (6.5, 0) {Transcribed Speech\\\& Timestamps};
\node[tbox, text width=3.5cm] (t_enc) at (6.5, -1.8) {Instruction\\Tokenizer};
\node[token] (t_q) at (6.5, -4.8) {$T_{L_q}$};

\node[concatbox] (concat) at (0, -6.0) {Multimodal Concatenation: $[T_V ; T_A^{global} ; T_A^1 ; \dots ; T_A^N ; T_{L_q}]$};
\node[llmbox] (llm) at (0, -7.5) {Large Language Model (LLaMA-2) + LoRA Fine-tuning};
\node[token, text=blue!70!black] (out) at (0, -8.8) {Emotion Prediction: ['anxious', 'concerned']};

\draw[->, thick] (v_in) -- (v_enc);
\draw[->, thick] (v_enc) -- (v_qf);
\draw[->, thick] (v_qf) -- (t_v);
\draw[->, thick] (t_v) -- (t_v |- concat.north);

\draw[->, thick] (a_in) -- (a_slide);
\draw[->, thick] (a_slide.south) +(-2.8,0) -- (a_glob.north);
\draw[->, thick] (a_slide.south) +(2.8,0) -- (a_loc.north);
\draw[->, thick] (a_glob) -- (t_a_g);
\draw[->, thick] (a_loc) -- (t_a_l);
\draw[->, thick] (t_a_g) -- (t_a_g |- concat.north);
\draw[->, thick] (t_a_l) -- (t_a_l |- concat.north);

\draw[->, thick] (t_in) -- (t_enc);
\draw[->, thick] (t_enc) -- (t_q);
\draw[->, thick] (t_q) -- (t_q |- concat.north);

\draw[->, thick] (concat) -- (llm);
\draw[->, thick] (llm) -- (out);

\end{tikzpicture}
}
\caption{The overall architecture of AcoustEmo. The multimodal pathways are strictly decoupled before late fusion. Crucially, the Utterance-Aware Acoustic Q-Former (center right) extracts local acoustic dynamics bounded by text timestamps, bypassing the limitations of purely global audio aggregation.}
\label{fig:architecture}
\end{figure*}

\section{Related Work}
\subsection{Multimodal Large Language Models}
The rapid advancement of Large Language Models (LLMs) has catalyzed the development of MLLMs designed for rich multimedia understanding. Models such as VideoChat \cite{videochat} and Video-LLaMA \cite{Video-llama} utilize Query Transformers (Q-Formers) \cite{blip2} to align visual and acoustic features with the LLM's text embedding space. Recently, audio-centric foundation models like Qwen-Audio \cite{Qwen-audio} and SALMONN \cite{salmonn} have further advanced the integration of speech and non-speech sounds into LLMs. However, these architectures generally process audio inputs as a single continuous stream or compress them into macroscopic representations. Traditional Speech Emotion Recognition (SER) systems often rely on handcrafted low-level descriptors (LLDs) such as pitch, jitter, and shimmer to capture transient prosodic traits. Unfortunately, integrating such high-resolution, uncompressed acoustic frames directly into LLMs remains computationally prohibitive, forcing a compromise where critical acoustic details are sacrificed for sequence compression.

\subsection{Emotion Recognition in MLLMs}
Speech Emotion Recognition traditionally relies on specialized acoustic models. With the advent of the EMER task \cite{EMER(Multi)}, the focus has shifted toward explainable, open-vocabulary reasoning using MLLMs. Methods like AffectGPT \cite{lian2024affectgpt} fine-tune LLMs to generate emotional descriptors based on multimodal inputs. Furthermore, Zhang et al. \cite{zhang2024microemo} highlighted the importance of subtle clue dynamics by introducing a global-local visual attention mechanism. Despite these efforts, the transient acoustic signals---such as breathiness or pitch micro-variations occurring within a fraction of a second during a specific spoken phrase---are often smoothed out by global pooling mechanisms. Our proposed AcoustEmo directly addresses this gap by imposing a structurally constrained, utterance-level acoustic modeling mechanism.

\section{Methodology}
We present AcoustEmo, a time-sensitive MLLM featuring a novel Utterance-Aware Acoustic Q-Former. 

\subsection{Overall Architecture}
As illustrated in Figure \ref{fig:architecture}, AcoustEmo processes multimodal inputs comprising video frames, audio tracks, and transcribed speech with timestamp boundaries. For the visual modality, we extract visual tokens $T_V$ using a pre-trained Vision Transformer and an Image Q-Former. The core innovation lies in the acoustic branch. Instead of generating a single global representation, the input audio is processed by our Utterance-Aware Acoustic Q-Former to generate multi-scale acoustic tokens $T_A$. Finally, $T_V$, $T_A$, and the instruction tokens $T_{L_{q}}$ are concatenated and fed into the Large Language Model to generate open-vocabulary emotion responses.

\subsection{Timestamp-Synchronized Sliding Window}
To model temporal acoustic relationships at a granular level, we propose a dynamic sliding window synchronized with utterance-level timestamps. Let the raw audio input be processed by a pre-trained acoustic encoder (e.g., ImageBind \cite{imagebind}) to obtain the frame-level acoustic feature sequence $F_A \in \mathbb{R}^{L \times d}$, where $L$ is the sequence length and $d$ is the feature dimension. 

Given the transcription timestamps for $N$ utterances in the dialogue, denoted as $\{(t_{start}^i, t_{end}^i)\}_{i=1}^N$, we design an utterance-aware sliding window. For the $i$-th utterance, we map the continuous time boundaries to the discrete feature sequence indices, extracting the corresponding local acoustic feature segment $F_A^i$:
\begin{equation}
    F_A^i = F_A[\lfloor t_{start}^i \cdot f_s \rfloor : \lceil t_{end}^i \cdot f_s \rceil]
\end{equation}
where $f_s$ represents the frame sampling rate of the acoustic encoder. By explicitly grounding the acoustic feature extraction to the linguistic boundaries, we ensure semantic coherence between modalities, preventing the unnatural truncation of words common in fixed-length windowing.

\subsection{Utterance-Aware Acoustic Q-Former}
Within each dynamic window, we implement an Acoustic Q-Former to compress and refine the local acoustic features. In our configuration, we initialize a set of learnable queries $Q \in \mathbb{R}^{K \times d_{model}}$, where $K=32$ represents the number of query tokens, and $d_{model}=768$ is the hidden dimension size. These queries interact with the local utterance features $F_A^i$ via cross-attention. Specifically, the queries are updated by attending to the acoustic frames:
\begin{equation}
    T_A^i = \text{Softmax}\left(\frac{(Q W_Q)(F_A^i W_K)^T}{\sqrt{d_k}}\right)(F_A^i W_V)
\end{equation}
where $W_Q, W_K, W_V$ are learned projection matrices. The cross-attention mechanism allows the $K$ queries to act as an information bottleneck, forcing the module to distil only the most emotionally salient micro-prosodic details for the $i$-th utterance. 

This process is repeated for all $N$ segments. Concurrently, a global acoustic token $T_{A}^{global}$ is extracted from the entire feature $F_A$ using a parallel global Q-Former to maintain the macroscopic background context. The final multi-scale fused acoustic tokens $T_A$ are obtained by concatenating the global and utterance-aware tokens:
\begin{equation}
    T_A = [T_{A}^{global}; T_A^1; T_A^2; \dots; T_A^N]
\end{equation}

\subsection{Multimodal Alignment and LLM Reasoning}
The combined sequence $[T_V; T_A; T_{L_{q}}]$ is fed into the LLM. To maximize the effectiveness of our temporal features, we design a specific, instruction-aware prompt template $T_{L_{q}}$. It explicitly incorporates the timestamp information corresponding to each utterance, compelling the LLM to align the textual context with the dynamically extracted acoustic tokens. The model is optimized using the standard causal language modeling loss:
\begin{equation}
    \mathcal{L} = -\sum_{t} \log P_{\Theta}(T_{L_{a}, t} | T_V, T_A, T_{L_{q}}, T_{L_{a}, <t})
\end{equation}
We apply Low-Rank Adaptation (LoRA) \cite{lora} to fine-tune the LLM efficiently. By freezing the vast majority of the foundation model's parameters, LoRA allows us to inject emotion-specific multimodal reasoning capabilities without suffering from the catastrophic forgetting of general world knowledge.

\section{Experiments}

\subsection{Experimental Setup}
\textbf{Datasets and Metrics}: We evaluate our model on the Explainable Multimodal Emotion Recognition (EMER) task \cite{emer_dataset}. Specifically, we utilize the test-set of EMER-Fine, a rigorous benchmark that provides rich, multi-faceted annotations including facial expressions, vocal tones, and semantic context across diverse video dialogues. The dataset encompasses not only Ekman's six basic emotions but also a wide spectrum of complex, nuanced affective states \cite{SimLabel, SimLabel1, Metric, QuMAB1, QuMAB, QuMATL}. The primary evaluation metrics are the newly established Accuracy and Recall metrics for the EMER task, evaluated based on the semantic similarity of the generated open-vocabulary lists compared to ground truth labels.

\textbf{Implementation Details}: We utilize LLaMA-2 (7B) \cite{llama2} as our foundation language model. The visual encoder and acoustic encoder parameters are kept completely frozen to preserve their zero-shot representation capabilities. We optimize the model using the AdamW optimizer with a base learning rate of 2e-5, $\beta_1=0.9$, $\beta_2=0.999$, and a weight decay of 0.05. A linear learning rate warmup is applied for the first 5\% of training steps, followed by a cosine decay schedule. The maximum sequence length for the LLM is set to 1024 tokens to comfortably accommodate the concatenated multimodal sequence and the generated text. We exclusively fine-tune the proposed Utterance-Aware Acoustic Q-Former, the projection layers, and the LoRA parameters for 3 epochs on a single NVIDIA GPU.

\subsection{Baseline Models}
To rigorously evaluate AcoustEmo, we benchmark it against three dominant categories of state-of-the-art architectures:
1) \textbf{Audio-centric LLMs} (e.g., Qwen-Audio \cite{Qwen-audio}, SALMONN \cite{salmonn}): These models exhibit exceptional generalized audio understanding but fundamentally lack visual grounding.
2) \textbf{Video-centric MLLMs} (e.g., Video-LLaMA \cite{Video-llama}, VideoChat2 \cite{VideoChat2}): These models integrate visual and acoustic streams but heavily rely on macroscopic temporal pooling, often losing fine-grained audio fidelity.
3) \textbf{Emotion-specific pipelines} (e.g., AffectGPT \cite{lian2024affectgpt}, MicroEmo \cite{zhang2024microemo, MicroEmo-arxiv}): These represent the current vanguard in multimodal emotion reasoning, with MicroEmo specifically emphasizing local visual dynamics.

\subsection{Quantitative Results}
Table \ref{tab:main_results} presents the comparison between AcoustEmo and the aforementioned state-of-the-art MLLMs. 

\begin{table}[h]
\caption{Main results on the test-set of EMER-Fine \cite{EMER(Multi)}.}
\label{tab:main_results}
\begin{tabular}{l@{\hspace{2.1mm}}c@{\hspace{2.1mm}}c@{\hspace{2.1mm}}c}
\toprule
Model & Avg & Accuracy$_S$ & Recall$_S$ \\
\midrule
\multicolumn{4}{c}{Audio + Subtitle} \\
\midrule
Qwen-Audio \cite{Qwen-audio} & 38.66 & 46.97 & 30.35 \\
OneLLM \cite{Onellm} & 40.56 & 42.55 & 38.56 \\
SECap \cite{Secap} & 52.78 & 61.36 & 44.19 \\
SALMONN \cite{salmonn} & 51.28 & 54.17 & 48.38 \\
\midrule
\multicolumn{4}{c}{Video + Subtitle} \\
\midrule
Otter \cite{Otter} & 37.72 & 42.22 & 33.22 \\
VideoChat \cite{videochat} & 46.34 & 41.49 & 51.19 \\
Video-LLaMA \cite{Video-llama} & 40.97 & 39.44 & 42.50 \\
Video-LLaVA \cite{Video-llava} & 42.75 & 45.38 & 40.13 \\
VideoChat2 \cite{VideoChat2} & 43.83 & 49.24 & 38.42 \\
OneLLM \cite{Onellm} & 53.40 & 58.65 & 48.14 \\
LLaMA-VID \cite{Llama-vid} & 50.90 & 51.69 & 50.11 \\
mPLUG-Owl \cite{mplug-owl} & 48.84 & 48.33 & 49.34 \\
Video-ChatGPT \cite{Video-chatgpt} & 46.12 & 50.00 & 42.25 \\
Chat-UniVi \cite{Chat-univi} & 53.20 & 54.29 & 52.11 \\
\midrule
\multicolumn{4}{c}{Audio + Video + Subtitle} \\
\midrule
SECap + mPLUG-Owl & 64.42 & 57.95 & 70.90 \\
SALMONN + Video-ChatGPT & 59.71 & 53.48 & 65.93 \\
SECap + Video-ChatGPT & 58.43 & 51.60 & 65.26 \\
SECap + Chat-UniVi & 60.38 & 51.39 & 69.37 \\
SALMONN + mPLUG-Owl & 65.15 & 55.28 & \bf{75.03} \\
SALMONN + Chat-UniVi & 62.64 & 55.84 & 69.44 \\
AffectGPT \cite{lian2024affectgpt} & 61.75 & 62.03 & 61.46 \\
MicroEmo \cite{zhang2024microemo} & 66.21 & 63.82 & 68.59 \\
AcoustEmo (Ours)  & \textbf{67.55} & \textbf{65.40} & 70.15 \\
\midrule
EMER (Multi) \cite{EMER(Multi)} & 79.31 & 80.91 & 77.70 \\
\bottomrule
\end{tabular}
\end{table}

\textbf{Discussion}: Analyzing the results in Table \ref{tab:main_results}, audio-centric models such as Qwen-Audio struggle due to the lack of visual context, highlighting the highly multimodal nature of the EMER task. Meanwhile, video-centric models like Video-LLaMA, despite integrating audio, fall short of our performance. This discrepancy stems from their reliance on global audio pooling, which invariably washes out the transient paralinguistic cues essential for complex emotion reasoning. In contrast, AcoustEmo establishes a fine-grained correspondence between acoustic features and linguistic content, yielding substantial gains (+5.80\% in Avg over AffectGPT) and outperforming the visually-focused MicroEmo \cite{zhang2024microemo} approach.

\subsection{Ablation Study}
To thoroughly validate the effectiveness of our core architectural contributions, we conducted a comprehensive ablation study by systematically removing or altering key modules of AcoustEmo. The comparative results are detailed in Table \ref{tab:ablation}.

\begin{table}[h]
  \caption{Ablation study of different architectural components on the EMER-Fine test set. We systematically evaluate the impact of removing the Global Acoustic Q-Former, replacing the timestamp-synchronized sliding window with a naive fixed-length (2s) window, and completely removing the Utterance-Aware Acoustic Q-Former.}
  \label{tab:ablation}
  \centering
  \begin{tabular}{lccc}
    \toprule
    \textbf{Model Settings}      & \textbf{Avg} & \textbf{Acc.} & \textbf{Rec.} \\
    \midrule
    \textbf{AcoustEmo (Full)}  & \textbf{67.55} & \textbf{65.40} & \textbf{70.15} \\
    \quad w/o Global Acoustic Q-Former & 64.10 & 63.25 & 65.15 \\
    \quad w/ fixed-length windows (2s) & 62.85 & 61.50 & 64.60 \\
    \quad w/o Utterance-Aware A-QF & 61.20 & 60.15 & 63.50 \\
    \bottomrule
  \end{tabular}
\end{table}

\textbf{Impact of Local Acoustic Dynamics:} The most drastic performance drop occurs when the Utterance-Aware Acoustic Q-Former is removed entirely. The average score plummets from 67.55 to 61.20. This highlights the absolute necessity of capturing local acoustic clues, as global averaging inherently dilutes instantaneous affective signals like sudden pitch shifts.

\textbf{Necessity of Timestamp Synchronization:} To verify the importance of strictly aligning audio segments with textual utterances, we replaced the timestamp-synchronized sliding window with a naive, fixed-length sliding window (e.g., extracting a local feature every 2 seconds). This variant suffers a substantial drop (Avg: 62.85). Fixed-length windows inevitably suffer from boundary mismatch, where a single window might arbitrarily slice a spoken word in half or encompass irrelevant silence between sentences. 

\textbf{Role of the Global Context:} Finally, we ablated the Global Acoustic Q-Former, relying solely on the local utterance-aware tokens. The performance decreases to an average of 64.10. This indicates that while local dynamics are paramount for capturing sudden emotional shifts, the holistic background context still provides valuable supplementary information for reasoning.

\subsection{Qualitative Analysis}
To further illustrate the advantage of AcoustEmo, we analyzed specific instances where the model succeeded while the global-feature baseline failed. In a video dialogue scenario, the speaker maintained a generally neutral visual expression and a steady speech rate for the majority of the utterance. However, in the final 1.5 seconds, there was a subtle vocal tremor (a micro-prosodic shift indicating anxiety).

\begin{figure}[h]
\centering
\resizebox{0.98\linewidth}{!}{
\begin{tikzpicture}[
    >=latex,
    box/.style={draw, rounded corners, align=center, font=\sffamily\footnotesize, inner sep=1.2ex},
    basebox/.style={box, fill=cyan!5, draw=cyan!60!black, text width=3.8cm},
    oursbox/.style={box, fill=orange!10, draw=orange!60!black, text width=3.8cm},
    predbox/.style={box, fill=gray!5, draw=gray!40, text width=3.6cm}
]

\node[anchor=south west, font=\sffamily\footnotesize] at (0, 0.8) {Subtitle: "It's fine, I just...};
\node[anchor=south west, font=\sffamily\footnotesize, text=red!70!black] at (4.5, 0.8) {[tremble] don't know what to do."};

\draw[->, thick, draw=black!70] (-0.2, -0.6) -- (7.8, -0.6) node[right, font=\sffamily\scriptsize] {Time};
\foreach \x/\label in {0/0.0s, 1.5/1.0s, 3.0/2.0s, 4.5/3.0s, 6.0/4.0s, 7.5/5.0s} {
    \draw (\x, -0.5) -- (\x, -0.7) node[below, font=\sffamily\scriptsize] {\label};
}

\foreach \x in {0, 0.05, ..., 5.25} {
    \draw[blue!60!black, thick] (\x, {-0.05 - 0.15*rnd}) -- (\x, {0.05 + 0.15*rnd});
}

\draw[fill=orange!20, draw=orange!80!black, thick, dashed] (5.25, -0.8) rectangle (7.5, 0.8);
\node[above, font=\sffamily\scriptsize\bfseries, text=orange!80!black, align=center] at (6.375, 1.3) {Voice Tremor\\(Local Anomaly)};
\foreach \x in {5.3, 5.35, ..., 7.45} {
    \draw[orange!90!black, thick] (\x, {-0.1 - 0.4*rnd}) -- (\x, {0.1 + 0.4*rnd});
}

\draw[->, thick, cyan!60!black, densely dotted] (3.75, -0.9) -- (1.8, -2.2);
\node[font=\sffamily\small\bfseries, fill=white, inner sep=3pt] at (1.8, -1.65) {A: Global Baseline};

\node[basebox] (base_enc) at (1.8, -2.8) {Global Audio Encoder\\(Averages entire 5s)};
\node[predbox] (base_pred) at (1.8, -4.2) {Prediction:\\\textcolor{cyan!70!black}{['calm', 'neutral']} \textcolor{red}{\textbf{X}}};
\draw[->, thick] (base_enc) -- (base_pred);

\draw[->, thick, orange!80!black] (6.375, -0.9) -- (5.8, -2.2);
\node[font=\sffamily\small\bfseries, fill=white, inner sep=3pt] at (5.8, -1.65) {B: AcoustEmo (Ours)};

\node[oursbox] (ours_enc) at (5.8, -2.8) {Utterance-Aware Q-Former\\(Isolates the anomaly)};
\node[predbox] (ours_pred) at (5.8, -4.2) {Prediction:\\\textcolor{orange!80!black}{['anxious', 'concerned']} \textcolor{green!60!black}{\textbf{\checkmark}}};
\draw[->, thick] (ours_enc) -- (ours_pred);

\end{tikzpicture}
}
\caption{Qualitative comparison between the global audio encoder and our AcoustEmo. The baseline averages out the transient micro-prosodic shift (voice tremor) at the end of the utterance, resulting in a neutral prediction. In contrast, AcoustEmo explicitly isolates this local acoustic anomaly via the utterance-aware sliding window, accurately deducing the underlying anxious state.}
\label{fig:qualitative}
\end{figure}
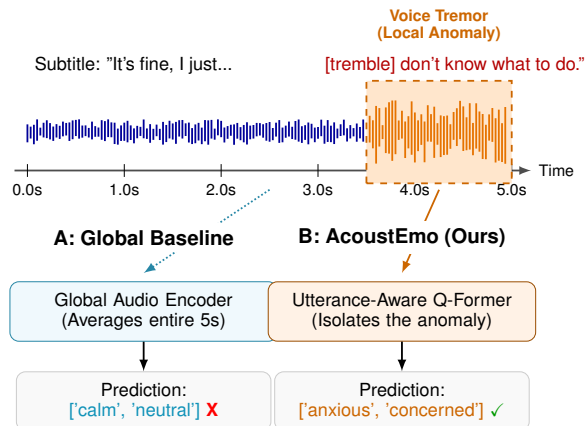

The baseline model, relying on the globally averaged acoustic tokens, completely smoothed out this brief anomaly, predicting the emotion simply as \textit{['calm', 'neutral']}. In contrast, AcoustEmo's sliding window mechanism successfully isolated the acoustic features of that specific timestamp boundary (Figure \ref{fig:qualitative}). The Utterance-Aware Acoustic Q-Former assigned higher cross-attention weights to the high-frequency fluctuations within that local window, enabling the LLM to accurately deduce the underlying emotional shift and output \textit{['anxious', 'concerned']}.

\subsection{Error Analysis}
Despite the significant improvements, AcoustEmo occasionally misclassifies highly ambiguous emotional states. For instance, in sarcastic utterances—where the acoustic tone directly contradicts the semantic meaning—the model can still be misled by the linguistic modality if the micro-prosodic cues are excessively subtle. Furthermore, background noise overlapping with the target speaker's voice boundaries can introduce noisy tokens into the Utterance-Aware Q-Former, slightly degrading performance in low-SNR (Signal-to-Noise Ratio) scenarios. Addressing these robust feature extraction challenges remains an avenue for future optimization.

\section{Conclusion}
In this paper, we introduced AcoustEmo, a time-sensitive Multimodal Large Language Model designed for open-vocabulary emotion reasoning. By proposing a novel Utterance-Aware Acoustic Q-Former, our framework successfully overcomes the limitations of traditional global audio encoders. The timestamp-synchronized sliding window explicitly captures local temporal dynamics and subtle acoustic clues, such as micro-prosody, within specific dialogue segments. Experimental results on the EMER task validate that deep acoustic modeling significantly enhances the model's capacity to trace evolving emotional states, establishing a more nuanced paradigm for speech-centric multimodal understanding. 

Future work will explore continuous emotion tracking by mapping these utterance-aware features directly into temporal arousal and valence spaces. Furthermore, we intend to investigate cross-lingual emotion recognition to evaluate the generalization capabilities of our architecture across diverse phonetic structures. Finally, optimizing the computational overhead of the dynamic segment extraction will be prioritized to facilitate real-time, low-latency emotion reasoning for on-device conversational agents, ultimately contributing to more empathetic and socially intelligent AI systems.


\bibliographystyle{IEEEtran}
\bibliography{mybib}

\end{document}